\documentclass[prd,showpacs,preprintnumbers]{revtex4}
\usepackage{amssymb,hyperref,pstricks}
\usepackage[dvips]{graphicx}

\def\Psibar{\overline{\Psi}}
\def\afi{\frac{1}{a_5}}
\def\gfive{(\gamma_5\otimes\openone)}
\def\gfivexfivemu{(\gamma_5\otimes\xi_{5\mu})}
\def\gonexfivemu{(\openone\otimes\xi_{5\mu})}
\def\gxone{(\openone\otimes\openone)}
\def\kp{k^\prime}
\def\meff{m_{\text{eff}}}
\def\qbar{\overline{q}}
\def\sp{s^\prime}
\def\xfive{(\openone\otimes\xi_5)}
\def\xfivemu{(\openone\otimes\xi_{5\mu})}
\def\xp{x^\prime}
\def\yp{y^\prime}

\begin{document}

\preprint{hep-lat/0207006}

\title{The staggered domain wall fermion method}

\author{George T.\ Fleming}
\email{gfleming@mps.ohio-state.edu}
\affiliation{Department of Physics, The Ohio State University, Columbus, OH
  43210, USA}

\author{Pavlos M.\ Vranas}
\email{vranasp@us.ibm.com}
\affiliation{IBM T.\ J.\ Watson Research Center, Yorktown Heights, NY
  10598, USA}

\date{October 3, 2002}

\begin{abstract}
A different lattice fermion method is introduced. Staggered domain wall
fermions are defined in $2n+1$ dimensions and describe $2^n$ flavors of
light lattice fermions with exact U(1)$\times$U(1) chiral symmetry in $2n$
dimensions.  As the size of the extra dimension becomes large, $2^n$ chiral
flavors with the same chiral charge are expected to be localized
on each boundary and the full SU($2^n$)$\times$SU($2^n$) flavor chiral symmetry
is expected to be recovered.  SDWF give a different perspective
into the inherent flavor mixing of lattice fermions and by design present
an advantage for numerical simulations of lattice QCD thermodynamics.
The chiral and topological index properties of the SDWF Dirac operator
are investigated.  And, there is a surprise ending...
\end{abstract}


\maketitle

\section{Introduction}
\label{sec:introduction}

Lattice fermions are elusive. They not only present enormous challenges
to numerical simulations of lattice QCD and other strongly interacting
field theories but also pose in a most blatant way the problem
of non-perturbative regularization of chiral gauge theories.
Obviously the two problems have a common source. In the past several years
enormous progress has been made in this direction. Interestingly,
extra dimensions have been used again in theoretical physics, except this time
the dimension is a tool to generate the correct low energy physics.

Domain wall fermions were introduced
in \cite{Kaplan:1992bt,Kaplan:1993sg,Narayanan:1993wx,Narayanan:1994sk,
Frolov:1993ck,Frolov:1994zr}.
A large volume of work has followed since. The reader is referred
to the annual reviews \cite{Kikukawa:2001jk,Edwards:2001ei,Vranas:2000tz,
Golterman:2000hr,Neuberger:1999ry,Luscher:1999mt,Blum:1998ud,Shamir:1996zx,
Creutz:1995px,Narayanan:1994gt} and references therein.
The DWF lattice regulator begins by defining a massive Wilson fermion
\cite{Wilson:1975id} in $2n+1$ dimensions.  If the boundary condition
at the edges of the extra dimension is free then chiral surface states develop
with the plus chirality fermion exponentially bound on one wall
and the minus chirality fermion on the other wall \cite{Shamir:1993zy}.
The two chiralities have an overlap that breaks chiral symmetry.
As the size $L_s$ of the extra dimension increases the overlap tends
to zero exponentially fast. As $L_s \to \infty$ the theory has
a single massless Dirac fermion in $2n$ dimensions.  Obviously,
this construction addresses both problems mentioned above.

Since the DWF construction starts with massive Wilson fermions, it is easy
to see that for any finite $L_s$ there can be none
of the exact chiral symmetries available in the naive lattice fermion
formulation.  In return for this shortcoming, flavor mixing between doubler
states inherent in naive fermions are pushed up to the scale
of the lattice cut-off, making them irrelevant.  Early in the development
of lattice field theories, the staggered approach to lattice fermions
was devised to preserve some of the exact chiral symmetries of naive fermions.
This meant there was still flavor mixing between the remaining doubler states
\cite{Kogut:1975ag,Banks:1976gq,Susskind:1977jm}.
In the spirit of staggered fermions, our domain wall construction presented
here will preserve some exact chiral symmetry at any $L_s$ at the cost
of introducing flavor mixing between light fermion doublers.
Flavor symmetry violations should be exponentially suppressed
in the $L_s \to \infty$ limit.  A preliminary version of this work
was presented in \cite{Fleming:2001ua}.

Staggered domain wall fermions (SDWF) are similar to DWF in the use
of exponential localization of surface states to counter unwanted features
of lattice regularization.  In our case, this means disentangling
the inherent flavor mixing between light doubler states.  As a result,
even at finite $L_s$, they have an exact U(1)$\times$U(1) chiral symmetry
very much like standard staggered fermions.  This property makes them
attractive for QCD thermodynamic simulations.  As $L_s \to \infty$
the light surface states in the theory are expected to recover
the full SU($2^n$)$\times$SU($2^n$) chiral symmetry.  It must be noted
that SDWF (and Wilson DWF as well) may not be able to develop light states
if the coupling is extremely strong.
Depending on one's perspective, SDWF combine the nice properties
of the domain wall method and staggered fermions.
For simulations of QCD thermodynamics with standard DWF the reader is referred
to \cite{Vranas:1999dg,Chen:2000zu,Fleming:2000bk,Fleming:2001qh}.

We would like to draw the attention of the reader to a more subtle issue
in this paper that might otherwise be overlooked.  In most of what follows,
the Saclay basis proposed by Kluberg-Stern \textit{et al.}\ and
predecessors \cite{Kluberg-Stern:1983dg,Gliozzi:1982,Doel:1983} is used
for its nice spin-flavor algebra.  While this is formally equivalent
to the standard bases used for numerical simulations
\cite{Golterman:1984cy,Daniel:1988aa}, the transformation is gauge dependent
and quite complicated.  However, the conclusions we draw
from our analytic work should be basis independent, The construction
of actions suitable for numerical simulation will be discussed at the end
of the paper.

One interesting result is that it may be possible to do numerical simulations
directly in the Saclay basis when Pauli-Villars fields are introduced.
This applies equally to staggered fermions as well as SDWF and may eliminate
a serious obstacle for using the Saclay basis for simulation.  The reason
is that the gauge field dependent transformation that accompanies
the basis change will cancel as part of the subtraction.  We believe
this is yet another example of the potential usefulness
of \textit{doubly regularized} lattice fermions \cite{Fleming:2000bk}.

For a semi-infinite extent in the extra dimension, the theory has four
chiral fermions with the same chiral charges and is anomalous.
To construct an anomaly free theory, we must use such ``quadruplets''
with charges as dictated by the corresponding anomaly cancellation condition.
This is completely analogous to the case of Wilson DWF.  Of course,
in order to simulate a two flavor theory, a dynamical algorithm
which effectively takes the square root of the fermionic determinant
should be used.  Further in the future, non-degenerate quark mass matrices
could be explored to simulate the four lightest quarks.

The paper is organized as follows: The SDWF Dirac operator and action
is defined in section~\ref{sec:sdwf}. The symmetries associated with the
SDWF action are presented in section~\ref{sec:symmetries}. The flavor content
is discussed in section~\ref{sec:flavors} and the free propagator is calculated
in section~\ref{sec:propagator}. The transfer matrix along the extra direction
is given in section~\ref{sec:tmatrix}. The promised surprise is
in section~\ref{sec:surprise} but the reader ought to work through
the preceding sections first...  The transcription to the single component
basis, problems and future directions are presented
in section~\ref{sec:single_comp}. A discussion about alternative actions
is given in section~\ref{sec:alt}. The paper is concluded
with section~\ref{sec:conclusions}.

\section{\label{sec:sdwf}Staggered Domain Wall Fermions}

In this section the SDWF Dirac operator and action is presented
in the Saclay basis \cite{Kluberg-Stern:1983dg}. Here, we show that
in the free theory light fermion fields localize exponentially
along the extra direction with suppressed flavor mixing.

The SDWF partition function is
\begin{equation}
Z = \int [dU] \int [d\Psibar d\Psi] \int [d\Phi^\dagger d\Phi]\ e^{-S} .
\label{eq:partition_func}
\end{equation}
$U_\mu(x)$ is the gauge field, $x$ is a site coordinate vector
in the $2n$ dimensional space and $\mu=1, 2, \cdots, 2n$.
$\Psi(y,s)$ is the fermion field and $\Phi(y,s)$ is a bosonic Pauli-Villars
(PV) field.  $y$ is a hypercube coordinate vector related to the site vector
$x$ by $x=2y + O + A$ where $O$ is a $2n$ dimensional binary vector indicating
one of the $2^{2n}$ possible origins of the hypercubic structure and $A$
is a binary vector which indicates position within the given hypercube $y$.
This implies the relations $y=(x - O - A)/2$ and $A=[(x - O) \bmod 2]$
between the vectors. $s=0,1,\cdots,L_s-1$ is a site coordinate
in the $2n+1$ direction, where $L_s$ is the number of sites
in this dimension.

The action $S$ is given by
\begin{eqnarray}
&&
S = S(\beta, L_s, m_0, m_f; U, \Psibar, \Psi, \Phi) = \\ \nonumber
&&
S_{\text{G}}(\beta; U) +
S_{\text{F}}(L_s, m_0, m_f; \Psibar, \Psi, V) +
S_{\text{PV}}(L_s, m_0; \Phi^\dagger, \Phi, V)
\label{eq:action}
\end{eqnarray}
where
\begin{equation}
S_{\text{G}} = \beta \sum_p \left[ 1 - \frac{1}{N_c} \text{Re\ Tr} U_p \right]
\label{eq:action_G}
\end{equation}
is the standard plaquette action with $\beta = 2 N_c / g_0^2$
with $g_0$ the lattice gauge coupling and $N_c$ the number of colors.
The fermion action is
\begin{equation}
S_{\text{F}} = - \sum_{y,\yp,s,\sp}
\Psibar(y,s) D_{\text{F}}(y,s; \yp,\sp) \Psi(\yp,\sp)
\label{eq:action_F}
\end{equation}
with the fermion matrix given by
\begin{equation}
D_{\text{F}}(y,s; \yp,\sp) = \delta(s-\sp) D(y,\yp)
  + D^\perp(s,\sp) \delta(y-\yp)
\label{eq:D_F}
\end{equation}
where $D(y,\yp)$ is the standard staggered action in the Saclay basis
with the typical staggered mass (distance zero) set to zero
and a different mass (distance one) proportional to $(1/a_5 - m_0)$
added as described below.  Here are the expressions in a chiral basis
in $2n=4$ dimensions.  Extensions to other even dimensions are straightforward.
\begin{equation}
D = 
\left( \begin{array}{cc}
   B         &  C     \\
  -C^\dagger & -B
\end{array} \right) ,
\label{eq:Dstag}
\end{equation}
\begin{equation}
B = - \sum_\mu \xfivemu \left[ \Delta_\mu(V)
+ \frac{m_0}{2} - \frac{1}{2 a_5} \right] ,
\label{eq:B}
\end{equation}
\begin{equation}
C = - \frac{1}{4} \sum_\mu \sigma_\mu \nabla_\mu(V) ,
\label{eq:C}
\end{equation}
\begin{equation}
\Delta_\mu(V; y, \yp) = \frac{1}{4} \left[ \delta(y+\hat\mu-\yp) V_\mu(y)
+ \delta(y-\hat\mu-\yp) V^\dagger_\mu(\yp) - 2 \delta(y-\yp) \right] ,
\label{eq:del_4}
\end{equation}
\begin{equation}
\nabla_\mu(V; y, \yp) = \frac{1}{4} \left[ \delta(y+\hat\mu-\yp) V_\mu(y)
- \delta(y-\hat\mu-\yp) V^\dagger_\mu(\yp) \right]
\label{eq:deriv_4}
\end{equation}
where $\sigma_{1,2,3}$ are the Pauli matrices and $\sigma_4$ is the identity.
$V_\mu(y)$ are the gauge links between hypercubes related to the $U_\mu(x)$
links of the gauge action by $V_\mu(y) = U_\mu(2y+O)\ U_\mu(2y+O+\hat\mu)$.
The parameter $m_0$ is the mass representing the ``\textit{height}''
of the domain wall.  The $s$-dependent part of the Dirac operator is
exactly as for DWF but with a different mass mixing. Here we consider
the action for $2^n$ degenerate flavors. For non-degenerate flavors the
action is similar and can be constructed according to the rules
outlined in section~\ref{sec:symmetries}.
\begin{equation}
D^\perp = D_5 + H(m_f) ,
\label{eq:D_perp_f}
\end{equation}
\begin{equation}
D_5(s, \sp) = \left\{ \begin{array}{ll} 
\afi \ P_R \delta(1-\sp)                              & s=0           \\ 
\afi [ P_R \delta(s+1-\sp) + P_L \delta(s-1-\sp)   ]  & 0 < s < L_s-1 \\ 
\afi \                       P_L \delta(L_s-2-\sp)    & s = L_s-1
\end{array}
\right. 
\label{eq:D_5}
\end{equation}
The mass mixing term depends to whether $L_s$ is even or odd.
For odd $L_s$ we have the following purely imaginary terms
\begin{equation}
H(m_f; s, \sp) = -\afi i\,m_f \left[
  P_R \delta(s-L_s+1) \delta(\sp) + P_L \delta(s) \delta(L_s-1-\sp)
\right]
\label{eq:H_5_odd}
\end{equation}
where $m_f$ is the degenerate mass of the flavor states localized
on the domain wall.  For even $L_s$ more care must be used in constructing
the mass mixing term to preserve the exact U(1)$\times$U(1) symmetry
(see section~\ref{sec:symmetries}).
In that case the mass term does not just involve the boundary at $s=0$
and $L_s-1$ but also at $s=1$ and $L_s-2$.  Furthermore notice that the terms
are real. The different ``reality'' of the mass term for $L_s$ even/odd
is just a reflection of a staggered wavefunction phase
of the form $i^{(s-\sp)}$. The even $L_s$ mass term is
\begin{eqnarray}
H(m_f; s, \sp) = - \frac{m_f}{a_5} & & \left[
    P_{LR} \delta(s)       \delta(L_s-2-\sp)
  + P_{LL} \delta(s-1)     \delta(L_s-1-\sp)
\right. \\
& & \left.
  + P_{RR} \delta(s-L_s+2) \delta(\sp)
  + P_{RL} \delta(s-L_s-1) \delta(1-\sp)
\right] . \nonumber
\end{eqnarray}
The chiral and flavor projectors are
\begin{equation}
P_{R,L} = P_\pm = \frac{ 1 \pm \gamma_5}{2}, \quad
F_{R,L} = F_\pm = \frac{ 1 \pm \xi_5}{2}, \quad
P_{RL,RL} = P_{\pm\pm} = (P_{\pm} \otimes F_{\pm}) .
\label{eq:projectors}
\end{equation}
The gamma matrices are taken in the standard chiral basis.  For example
$2n=4$ dimensions they are chosen to be
\begin{equation}
\gamma_\mu = \left( \begin{array}{cc} 0 & \sigma_\mu \\ \sigma_\mu^\dagger & 0 \end{array} \right), 
\mu=1,2,3\ \  
\gamma_4 = \left( \begin{array}{cc} 0 & 1 \\ 1 & 0 \end{array} \right), \ \ 
\gamma_5 = \left( \begin{array}{cc} 1 & 0 \\ 0 & -1 \end{array} \right) 
\label{eq:gamma}
\end{equation}
with $\sigma_\mu$ the Pauli matrices. The flavor matrices are
defined as usual \cite{Kluberg-Stern:1983dg}
\begin{equation}
\xi_\mu = \gamma_\mu^T
\label{eq:xi}
\end{equation}
and notations like $\gfivexfivemu$ mean $(\gamma_5\otimes\xi_5\xi_\mu)$.

As with DWF, the PV action is designed to cancel the contribution
of the heavy fermions.  This is necessary because the number of heavy fermions
is $\sim L_s$ and in the $L_s \rightarrow \infty$ limit they produce bulk type
infinities \cite{Narayanan:1993wx,Narayanan:1994sk,Narayanan:1993ss,
Narayanan:1995gw}. There is some flexibility in the definition of the PV action
since different actions could have the same $L_s \rightarrow \infty$ limit.
However, the choice of the PV action may affect the approach to the $L_s
\rightarrow \infty$ limit. Here the same approach
as in \cite{Vranas:1997tj,Vranas:1998da} was chosen. The $m_f=1$ case
is exactly  the quenched theory (infinitely massive fermions).
The PV action is
\begin{equation}
S_{\text{PV}} = \sum_{x,\xp,s,\sp}
  \Phi^\dagger(x,s)D_{\text{F}}[m_f=1](x,s; \xp, \sp) \Phi(\xp,\sp) .
\label{eq:action_PV}
\end{equation}

The symmetries and detailed properties of the SDWF Dirac operator
will be discussed in the rest of the paper.
However, as a first check we verify that in the free case 
the SDWF Dirac operator indeed describes four flavors with the chiralities
localized on the opposite walls. Following identical steps
as in \cite{Kaplan:1992bt} we go to momentum space and demands that
in order for light modes to exist there must be a wavefunction such that
\begin{equation}
D_{\text{F}}(k,s; \kp, \sp) \phi(\kp,\sp)
  = D_{\text{naive}}(k,s; \kp, \sp) \phi(\kp,\sp) .
\label{eq:free_loc_cond_1}
\end{equation}
In essence this equation demands that the extra term in the Dirac operator
$D^\perp$ cancels the flavor breaking term $B$. From the above equations it 
is easy to see that Eq.~(\ref{eq:free_loc_cond_1}) leads to
\begin{equation}
\label{eq:zero_mode_problem}
\sum_{\sp} \left\{ \textstyle\frac{1}{a_5} \left[
P_+ \delta(s+1-\sp) + P_- \delta(s-1-\sp) \right]
+ {\textstyle \sum_\mu} \gfivexfivemu b_\mu \delta(s-\sp)
\right\} \phi(k,\sp) = 0
\end{equation}
where
\begin{equation}
b_\mu = \frac{ 1 - \cos{k_\mu} - m_0 + 1/a_5}{2} .
\label{eq:b_mu}
\end{equation}
From this equation, it is easy to see that the $P_\pm$ projectors
in the $s$-dependent part commute with the flavor breaking part
so that each may be simultaneously diagonalized.  This constraint alone
effectively restricts the allowed projectors to the ones chosen here.

\begin{figure}
\includegraphics[width=0.9\columnwidth]{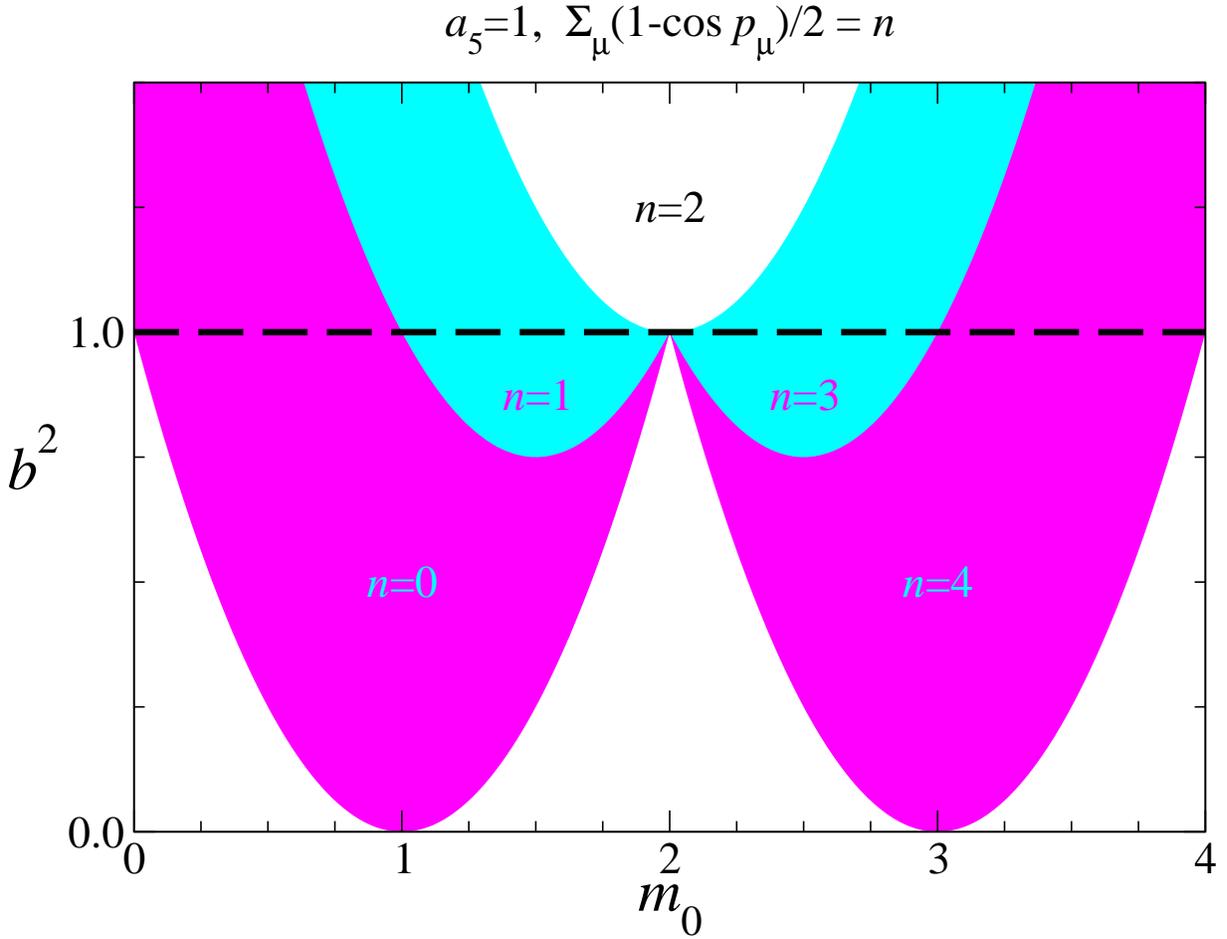}
\caption{\label{fig:m0_range}From Eq.~(\ref{eq:free_loc_cond_3}): $b^2$
  at the origins of the five Brillouin zones $n=[0,1,2,3,4]$ plotted
  \textit{vs.}\ $m_0$.}
\end{figure}

The solution is separable and $\phi(s)$ is the $s$-dependent part
\begin{equation}
\label{eq:block_vector}
\phi(s) = \left(
\phi_{++}, \phi_{+-}, \phi_{-+}, \phi_{--} 
\right)
\end{equation}
where $\phi_{-+}(s) = P_{-+} \phi(s)$, \textit{etc}.  In this notation,
we can write
\begin{equation}
\label{eq:B_matrix}
{\textstyle \sum_\mu} (\gamma_5\otimes\xi_{5\mu}) b_\mu = \left(
  \begin{array}{rrrr}
                        & \overline{b} &                      &               \\
  -\overline{b}^\dagger &              &                      &               \\
                        &              &                      & -\overline{b} \\
                        &              & \overline{b}^\dagger &
  \end{array}
\right)
\end{equation}
where
\begin{equation}
{\overline{b}_j} = i \sigma_j^{*} b_j, \quad j=1,2,3,
\quad {\overline{b}_4} = b_4 .
\label{eq:b_bar}
\end{equation}
Solving Eqs.~(\ref{eq:zero_mode_problem}) relating nearest neighbor $s$ sites
is a bit complicated because the flavor components mix and is not of interest
for this discussion.  On the other hand the solutions to these equations
after iterating twice are simple.  For $a_5=1$ we have 
\begin{equation}\begin{array}{rcl}
\label{eq:zero_mode_solution}
\phi_{\pm+}(s\pm2) & = & -\overline{b} \overline{b}^\dagger \phi_{\pm+}(s) , \\
\phi_{\pm-}(s\pm2) & = & -\overline{b}^\dagger \overline{b} \phi_{\pm-}(s) .
\end{array}\end{equation}
For free fermions, $[\overline{b},\overline{b}^\dagger]=0$
and $\overline{b} \overline{b}^\dagger$, $\overline{b}^\dagger \overline{b}$
are both proportional to the identity with eigenvalue
\begin{equation}
\lambda(\overline{b} \overline{b}^\dagger)
= \lambda(\overline{b}^\dagger \overline{b}) = b^2 ,
\label{eq:free_eig}
\end{equation}
\begin{equation}
b = \sqrt{{\textstyle \sum_\mu} b_\mu^2} .
\label{eq:free_b}
\end{equation}
If we require that
\begin{equation}
b^2 < 1
\label{eq:free_loc_cond_2}
\end{equation}
then for a semi-infinite $s$ direction, $s$$\ge$0, only $\phi_{+\pm}$
is normalizable, while $\phi_{-\pm}$ is not. However, this is not
enough to ensure that the doubler modes are not present.  We must
further require that the above condition excludes momenta with
components larger or equal to $\pi$. This can be seen by writing out 
Eq.~(\ref{eq:free_loc_cond_2})
\begin{equation}
b^2 = \frac{1}{4} \sum_\mu \left[ (1 - \cos{k_\mu}) + (1 - m_0) \right]^2 < 1 .
\label{eq:free_loc_cond_3}
\end{equation}
For momenta near the origins of the $n^{\text{th}}$ Brillouin zone
(where $n$ is the number of momentum components near $\pi$) this gives
\begin{equation}
(1 - m_0)^2 + n (1 - m_0) + n < 1 .
\label{eq:free_loc_cond_4}
\end{equation}
When $n=0$ the sufficient condition is
\begin{equation}
0 < m_0 < 2 ,
\label{eq:free_loc_cond_5}
\end{equation}
same as for Wilson DWF. However, this condition does not ensure that
the doubler modes are non-normalizable for all $n$.  For example, for $n=1$
and $m_0 = 1.5$ both Eqs.~(\ref{eq:free_loc_cond_4})
and (\ref{eq:free_loc_cond_5}) are satisfied making the $1^{\text{st}}$
Brillouin zone doubler wave functions normalizable.  The range of $m_0$
needs to be further restricted. The following condition ensures that
only the $0^{\text{th}}$ Brillouin zone wave function is normalizable
\begin{equation}
0 < m_0 < 1 .
\label{eq:free_loc_cond_6}
\end{equation}
The above is presented graphically in Fig.~\ref{fig:m0_range}.
It is straightforward to extend these results to the more general case
of $0 < a_5 < 1$.

\section{Symmetries}
\label{sec:symmetries}

When constructing the SDWF action, it is important to preserve
the symmetries of the massless staggered action
\cite{Kogut:1975ag,Jolicoeur:1986ek}.
Of course, adding any new terms to the staggered action will break some
of those symmetries, so we have to find new symmetries that involve
the extra dimension. The symmetry transformations for the action in
the Saclay basis of section \ref{sec:sdwf} are presented below.

U(1)$_e$$\times$U(1)$_o$ \textit{chiral rotations}.  
The presence of this symmetry is one of the main motivations of this paper.
The residual chiral symmetry of staggered fermions involves making 
separate chiral rotations on even and
odd sites.  Terms in $D^\perp$ are not invariant under these rotations unless
we extend the notion of even and odd, \textit{including the extra dimension}.
The operator $\mathcal{S}$ is defined as
\begin{equation}
\mathcal{S}_{s,\sp} \equiv (-1)^s  \delta(s - \sp)
\label{eq:s_op}
\end{equation}
and then the extended even/odd projection operators are defined as
\begin{eqnarray}
\overline{P}_e & = & {\textstyle \frac{1}{2}} \left[
  (\openone\otimes\openone) + \mathcal{S} (\gamma_5\otimes\xi_5)
\right] , \nonumber \\*
\overline{P}_o & = & {\textstyle \frac{1}{2}} \left[
  (\openone\otimes\openone) - \mathcal{S} (\gamma_5\otimes\xi_5)
\right] .
\label{eq:eo_proj}
\end{eqnarray}
Using these projection operators the chiral transformation is
\begin{eqnarray}
\Psi(y) & \to & \left( e^{i\theta_e} \overline{P}_e 
+ e^{i\theta_o} \overline{P}_o \right)\ \Psi(y) ,
\nonumber \\*
\Psibar(y) & \to & \Psibar(y)\ 
\left( e^{-i\theta_o} \overline{P}_e + e^{i\theta_e} \overline{P}_o \right) .
\label{eq:chiral_trans}
\end{eqnarray}

\textit{Rotations by $\pi/2$}. These rotations are in planes perpendicular
to the extra dimension and the transformations are the same as the original
staggered ones.

\textit{$\mu$-parity}.  These transformations reflect the $2n-1$ spacetime axes
perpendicular to the spacetime axis in the $\hat\mu$ direction.  $D^\perp$
is not invariant under this symmetry unless we also reflect the $s$ direction
as well.  If the reflection operator is defined as
\begin{equation} 
\mathcal{R}_{s,\sp} \equiv \delta(L_s - 1 - s - \sp)
\label{eq:mu_parity_op}
\end{equation}
then the transformation is
\begin{equation} 
\begin{array}{rcl}
\Psi(y,s) & \to &
  (\gamma_\mu\otimes\xi_5) \mathcal{R}_{s,\sp}\ \Psi(y,\sp) , \\*
\Psibar(y,s) & \to &
  \Psibar(y,\sp)\ \mathcal{R}_{\sp,s} (\gamma_\mu\otimes\xi_5) .
\end{array} 
\label{eq:mu_parity_trans}
\end{equation}

\textit{Shift by one lattice spacing}.
The $m_0 - 1/a_5$ term in the SDWF action breaks this standard
staggered fermion symmetry at the expense of absorbing the renormalization
of the flavor breaking term.  That these terms are additively renormalized
in the interacting theory follows from the work of Mitra and Weisz
\cite{Mitra:1983bi}.  Nevertheless, interesting methods to alleviate
the breaking of this symmetry are discussed in section~\ref{sec:single_comp}.
This symmetry relates to interactions inside a hypercube which are essentially
non-physical.  We feel that the breaking of this symmetry is a small sacrifice,
but the issue certainly can and should be debated.

The symmetry transformation for the case $m_0 = 1/a_5$ is given.
Already for staggered fermions, the symmetry transformation is complicated
in the Saclay basis due to the imposed hypercubic structure of the formulation.
For SDWF, there is an added complication. Some parts of the transformation
require a reflection in the $s$ direction
\begin{eqnarray}
\lefteqn{ \Psi(y) \to {\textstyle \frac{1}{2}} \left[
  (\openone\otimes\xi_\mu) - (\gamma_{\mu5}\otimes\xi_5) \mathcal{R}
\right] \Psi(y) } \nonumber \\*
& & + {\textstyle \frac{1}{2}} \left[
  (\openone\otimes\xi_\mu) + (\gamma_{\mu5}\otimes\xi_5) \mathcal{R}
\right] \Psi(y+\hat\mu) , \nonumber \\*
\lefteqn{ \Psibar(y) \to \Psibar(y) {\textstyle \frac{1}{2}}
\left[
  (\openone\otimes\xi_\mu) - \mathcal{R} (\gamma_{5\mu}\otimes\xi_5)
\right] } \nonumber \\*
& & + \Psibar(y+\hat\mu) {\textstyle \frac{1}{2}} \left[
  (\openone\otimes\xi_\mu) + \mathcal{R} (\gamma_{5\mu}\otimes\xi_5)
\right]
\label{eq:one_shift_trans}
\end{eqnarray}
where $s$ indices have been suppressed.

\section{Flavors of SDWF}
\label{sec:flavors}

In this section the SDWF flavor identification is made in the Saclay basis.
From section~\ref{sec:sdwf} we see that for a finite extra direction
with $L_s$ sites the $P_+$ components of all flavors are localized
around $s$=0 while the $P_-$ components are localized around
$s$=$L_s$$-$1.  However, as already mentioned in section~\ref{sec:symmetries}
one of the main goals of this paper is to preserve most
of the staggered symmetries and particularly the U(1)$_e$$\times$U(1)$_o$
chiral symmetry.  For example, to generate the four-dimensional
flavor components $q$ with $P_{++}$, we should choose $s$ near zero.
 If $s$=0 is chosen, $P_{++} q(y) = P_{++} \Psi(y,0)$, then these components
also belong to the $P_e$ part of the fermion field.  Therefore, to
project flavor components with $P_{-+}$ we are not only restricted
to choose $s$ near $L_s$$-$1 but also choose $s$ so these components
belong to the $P_o$ part of the fermion field.  Then, components
$P_{\pm+} q$ will not mix even for finite $L_s$ because of the
even/odd symmetry.  In this example, we would like to pick $P_{-+}
q(y) = P_{-+} \Psi(y,s)$ with $s$ being even and near $L_s$$-$1.  So, if
$L_s$ is odd then $s$=$L_s$$-$1 is a good choice.  However, if $L_s$
is even, then we should choose $s$=$L_s$$-$2 instead.
\begin{equation}
\textstyle
\begin{array}{c}
\left[ 
\begin{array}{c}
\left(
\begin{array}{c} 
\text{X} \\ 
x \\
x \\
x 
\end{array}
\right) \\ \\
\left(
\begin{array}{c} 
x \\ 
\text{X} \\
x \\
x 
\end{array}
\right) \\ \\
\vdots \\ \\
\left(
\begin{array}{c} 
x \\ 
x \\
\text{X} \\
x 
\end{array}
\right) \\ \\
\left(
\begin{array}{c} 
x \\ 
x \\
x \\
\text{X} 
\end{array}
\right)
\end{array}
\right] \\ \\
\Psi
\end{array}
\!\!\!\!=\!\!\!\!
\begin{array}{c}
\left[ 
\begin{array}{c}
\left(
\begin{array}{c} 
\text{X} \\ 
0 \\
0 \\
x 
\end{array}
\right) \\ \\
\left(
\begin{array}{c} 
0 \\ 
\text{X} \\
x \\
0 
\end{array}
\right) \\ \\
\vdots \\ \\
\left(
\begin{array}{c} 
x \\ 
0 \\
0 \\
x 
\end{array}
\right) \\ \\
\left(
\begin{array}{c} 
0 \\ 
x \\
x \\
0 
\end{array}
\right)
\end{array}
\right] \\ \\
P_e \Psi
\end{array}
\!\!\!\!+\!\!\!\!
\begin{array}{c}
\left[ 
\begin{array}{c}
\left(
\begin{array}{c} 
0 \\ 
x \\
x \\
0 
\end{array}
\right) \\ \\
\left(
\begin{array}{c} 
x \\ 
0 \\
0 \\
x 
\end{array}
\right) \\ \\
\vdots \\ \\
\left(
\begin{array}{c} 
0 \\ 
x \\
\text{X} \\
0 
\end{array}
\right) \\ \\
\left(
\begin{array}{c} 
x \\ 
0 \\
0 \\
\text{X} 
\end{array}
\right)
\end{array}
\right] . \\ \\
P_o \Psi
\end{array}
\label{eq:flavors_ls_even}
\end{equation}

Using the block notation of Eq.~(\ref{eq:block_vector}), an example
for even $L_s$ is sketched in Eq.~(\ref{eq:flavors_ls_even}).  In this equation
$\Psi$($s$=0) is at the top and $\Psi$($s$=$L_s$$-$1) is at the bottom.
The capital letters denote one of the correct choices.  On the other hand
if $L_s$ is odd, \textit{e.g.}\ $L_s$=3, then
\begin{equation}
\begin{array}{c}
\left[ 
\begin{array}{c}
\left(
\begin{array}{c} 
\text{X} \\ 
\text{X} \\
x \\
x 
\end{array}
\right) \\ \\
\left(
\begin{array}{c} 
x \\ 
x \\
x \\
x 
\end{array}
\right) \\ \\
\left(
\begin{array}{c} 
x \\ 
x \\
\text{X} \\
\text{X}
\end{array}
\right)
\end{array}
\right] \\ \\
\Psi
\end{array}
\!\!\!\!=\!\!\!\!
\begin{array}{c}
\left[ 
\begin{array}{c}
\left(
\begin{array}{c} 
\text{X} \\ 
0 \\
0 \\
x 
\end{array}
\right) \\ \\
\left(
\begin{array}{c} 
0 \\ 
x \\
x \\
0 
\end{array}
\right) \\ \\
\left(
\begin{array}{c} 
x \\ 
0 \\
0 \\
\text{X} 
\end{array}
\right)
\end{array}
\right] \\ \\
P_e \Psi
\end{array}
\!\!\!\!+\!\!\!\!
\begin{array}{c}
\left[ 
\begin{array}{c}
\left(
\begin{array}{c} 
0 \\ 
\text{X} \\
x \\
0 
\end{array}
\right) \\ \\
\left(
\begin{array}{c} 
x \\ 
0 \\
0 \\
x 
\end{array}
\right) \\ \\
\left(
\begin{array}{c} 
0 \\ 
x \\
\text{X} \\
0 
\end{array}
\right)
\end{array}
\right] . \\ \\
P_o \Psi
\end{array}
\label{flavors_ls_odd}
\end{equation}
We note that other choices for selecting flavor components near the boundaries
are certainly possible.

\section{The SDWF propagator}
\label{sec:propagator}
The propagator in the Saclay basis in momentum space and for $m_f=0$
has the form
\begin{equation}
D^{-1}(s,\sp) =G_1\ \epsilon(s - \sp) + G_3\ \epsilon(s - \sp - 1) .
\label{eq:prop_massless}
\end{equation}
For $m_f \ne 0$ and $L_s = \text{multiple of 4}$ has the form
\begin{equation}
D^{-1}(s,\sp) = \left[ G_1 + m_f G_2 \right]\ \epsilon(s - \sp)
+ G_3\ \epsilon(s - \sp - 1) .
\label{eq:prop_massive}
\end{equation}
For $L_s = \text{even but not multiple of 4}$, the propagator has the same form
as in Eq.~(\ref{eq:prop_massive}) but with $m_f \to -m_f$.
For $L_s = \text{odd}$ the $m_f$ term is more complicated
and is not given here.  Nevertheless, based on sections~\ref{sec:sdwf}
and \ref{sec:symmetries} one can see that the form is similar.

$G_1(p;s,\sp)$, $G_2(p;s,\sp)$ and $G_3(p;s,\sp)$ ($p$ is the momentum) 
are proportional to the identity in their flavor indices. Also, $G_1$
anti-commutes with $\gfive$ while $G_2$ and $G_3$ commute with $\gfive$.
The flavor mixing is in the function $\epsilon(s)$
\begin{equation} \begin{array}{rcll} \displaystyle
\epsilon(s) & = & \gxone , & (s\ \text{even}) \\*
\epsilon(s) & = &
\frac{\textstyle\sum_\mu \gonexfivemu b_\mu}{\textstyle b}, &
(s\ \text{odd})
\label{eq:epsilon}
\end{array} \end{equation}
where $|b|$ is given in Eq.~(\ref{eq:free_b}).  For $s-\sp$ even and $m_f$=0
the propagator has no flavor mixing except for the $G_3$ term
in Eq.~(\ref{eq:prop_massless}).  In this term $\epsilon(\text{odd})$ breaks
flavor in exactly the same way as for free staggered fermions.
An exact U(1)$\times$U(1) symmetry is maintained.  The matrix coefficient
$G_3$ vanishes exponentially fast with $L_s$ for $s$, $\sp$ near
opposing boundaries and therefore as $L_s \to \infty$ with $m_f$=0 the
propagator anti-commutes with $\gfive$ and has no flavor mixing
provided $s-\sp$ is even.  This is in accordance with the discussion
in section~\ref{sec:flavors}. For $s-\sp$ odd more severe flavor mixing 
is present.

The $G_1$, $G_2$ and $G_3$ terms are
\begin{equation} 
G_1(p:s,\sp) = i^{(s - \sp -1)} \frac{1}{2} \sum_\mu \gamma_\mu \sin{p_\mu}
\ \left[ P_+ G(s,\sp) + P_- G(L_s-1-s, L_s-1-\sp) \right] ,
\label{eq:prop_G_1}
\end{equation}
\begin{equation} 
G_2 = - i^{(s - \sp -1)} \
\ \left[ P_+ \delta(s, 0) G(L_s -1,\sp) 
- \delta(s, L_s-1) P_- G(L_s-1, L_s-1-\sp) \right] ,
\label{eq:prop_G_2}
\end{equation}
\begin{eqnarray} 
G_3 &=& i^{(s - \sp -1)} P_+ \left[
  b G(s,\sp) + i \theta_+(s) G(s-1, \sp)
\right] \\ \nonumber 
&+&  P_- \left[
  b G(L_s-1-s,L_s-1-\sp) + i \theta_-(s) G(L_s-2-s, L_s-1-\sp)
\right] .
\label{eq:prop_G_3}
\end{eqnarray}
where
\begin{equation} \begin{array}{rcll} \displaystyle
\theta_+(s) & = & 1 , & (s \ne 0) \\*
\theta_+(s) & = & 0 , & (s =   0) \\*
\theta_-(s) & = & 1 , & (s \ne L_s-1) \\*
\theta_-(s) & = & 0 , & (s =   L_s-1) .
\end{array} \label{eq:theta} \end{equation}
There is an exact correspondence with the terms of the standard DWF propagator.
In the notation of \cite{Vranas:1997tj,Vranas:1998da}
the symbol correspondences between SDWF and DWF are
\begin{equation} 
G(p:s,\sp) \to G_{+}(p:s,\sp), \ \ \ 
b \to b, \ \ \ 
\overline{p} \to \frac{1}{4} \overline{p} .
\label{eq:dwf_equiv}
\end{equation}
The reader is referred there for the detailed forms.

As can be seen the decay coefficient is now in terms of $b$ given 
in Eq.~(\ref{eq:free_b}) instead of the DWF 
$b = \sum_\mu [ 1 - \cos{p_\mu}] + 1/a_5 - m_0$. 
The localization condition for $m_0$ is over half the DWF range (see
Eq.~(\ref{eq:free_loc_cond_5})). Also, for $L_s$ odd, the effective
mass $\meff$ has the same general form as in Wilson DWF
\cite{Vranas:1997tj,Vranas:1998da}
\begin{equation}
\meff = (1 - \frac{2n}{4} m_0^2)(m_f + |1 - m_0|^{L_s}) .
\label{eq:meff}
\end{equation}

\section{The SDWF transfer matrix}
\label{sec:tmatrix}

The SDWF transfer matrix in the Saclay basis is presented.
We can use the technique of Neuberger \cite{Neuberger:1998bg}
to rewrite the free SDWF determinant in a form that allows
for a  quick identification of the transfer matrix.  A complete Hamiltonian
analysis is beyond the scope of this work. 

After interchanging various rows and columns of the SDWF matrix,
the determinant is equivalent to the determinant of the matrix
\begin{equation} \left( \begin{array}{cccc}
  \alpha_0 &        &                & \beta_0        \\
  \beta_1  & \ddots &                &                \\
           & \ddots & \alpha_{L_s-2} &                \\
           &        & \beta_{L_s-1}  & \alpha_{L_s-1}
\end{array} \right) 
\label{eq:sdwf_mat_suffled}
\end{equation}
where all of the $\alpha_s$ and $\beta_s$ are the block triangular matrices
\begin{equation}
\alpha_s = \left( \begin{array}{cc}
  -B  & 0     \\
   C  & 1/a_5
\end{array} \right) , \quad
\beta_s = \left( \begin{array}{cc}
  1/a_5 & -C^\dagger \\
  0     &  B
\end{array} \right) .
\label{eq:alpha_beta}
\end{equation}
For $\alpha_{L_s-1}$ and $\beta_0$, $1/a_5$ is replaced with $-\mu/a_5$
so $\mu$ is a parameter that controls the boundary conditions:
$\mu = \pm 1$ for (anti)periodic and $\mu=0$ for free. This parameter is 
used only for the derivation of the transfer matrix and it is not needed
in the theory.  In particular, the reader is cautioned against using $\mu$
as a mass parameter since it is inconsistent with the rules
of section~\ref{sec:flavors}.  The definitions of $B, C$ are given
in Eqs.~(\ref{eq:B}) and (\ref{eq:C}).

In this notation, following Neuberger's construction leads
to the SDWF determinant
\begin{equation}
D_{\text{F}} = (\text{fixed\ sign}) ( \det{B/a_5} )^{L_s}
\det{
\left[ \left( \begin{array}{cc}
  -\mu &  0 \\
   0   &  1
\end{array} \right) 
- T^{-L_s}
\left( \begin{array}{cc}
   1  &  0 \\
   0  & -\mu
\end{array} \right) \right]
}
\label{eq:sdwf_det}
\end{equation}
and the SDWF transfer matrix identification
\begin{equation}
T = - \left( \begin{array}{cc}
  B^{-1}/a_5        &  B^{-1} C \\
  C^\dagger B^{-1}  & a_5 [C^\dagger B^{-1} C - B]
\end{array} \right) .
\label{eq:transfer_mat}
\end{equation}
We can easily check that
\begin{equation}
[C, \xfive] = 0, \quad \{B, \xfive\} = 0
\label{eq:C_B_comm_anticom}
\end{equation}
and as a result
\begin{equation}
\{T, \xfive\} = 0 .
\label{eq:T_anticom}
\end{equation}
Also since $B^\dagger = - B$ we can see that $T$ is also anti-Hermitian
\begin{equation}
T^\dagger = - T .
\label{eq:T_antiherm}
\end{equation}
This is different from Wilson DWF and gives some idea why solving the
zero mode problem in Eq.~(\ref{eq:zero_mode_solution}) simplifies when
solving for the field two sites away. Obviously, standard transfer matrix 
manipulations  should be done with the Hermitian transfer matrix $T^2$ which
corresponds to a Hermitian Hamiltonian $H$.

\section{About that surprise...}
\label{sec:surprise}

At this point the reader must be wondering:  ``What is the spectrum of the 
transfer matrix?'' and ``What is the corresponding Hamiltonian?'' Here is 
where we were a bit surprised. Just in case the reader will later 
(after reading this section) be tempted to claim that there is no surprise,
she/he is invited to guess the $a_5 \to \infty$ limit Hamiltonian as well as 
the general $a_5$ spectrum of $T^2$.

We find, after some algebra, the following Hamiltonian $H_0$ corresponding
to the $a_5 \to 0$ limit of the transfer matrix
\begin{equation}
\lim_{a_5 \to 0} - T^2 = e^{-2 a_5 H_0}, \quad
H_0 = -\gfive \left( \begin{array}{cc}
  \frac{1}{2} \sum_\mu \Delta_\mu + m_0   &  C \\
  -C^\dagger                              &  \frac{1}{2} \sum_\mu \Delta_\mu + m_0
\end{array} \right) .
\label{eq:Hamiltonian}
\end{equation}
The first surprise is that $H_0$ is exactly diagonal in flavor.  It is almost,
but not exactly, the same as the standard overlap Hamiltonian $H_w$,
as $\Delta_\mu$ has a factor of $\frac{1}{2}$ compared to $C$.
Nevertheless, because $H_0$ is diagonal in flavor, the standard machinery
of DWF can be directly applied.

\begin{figure}
\includegraphics[width=0.9\columnwidth]{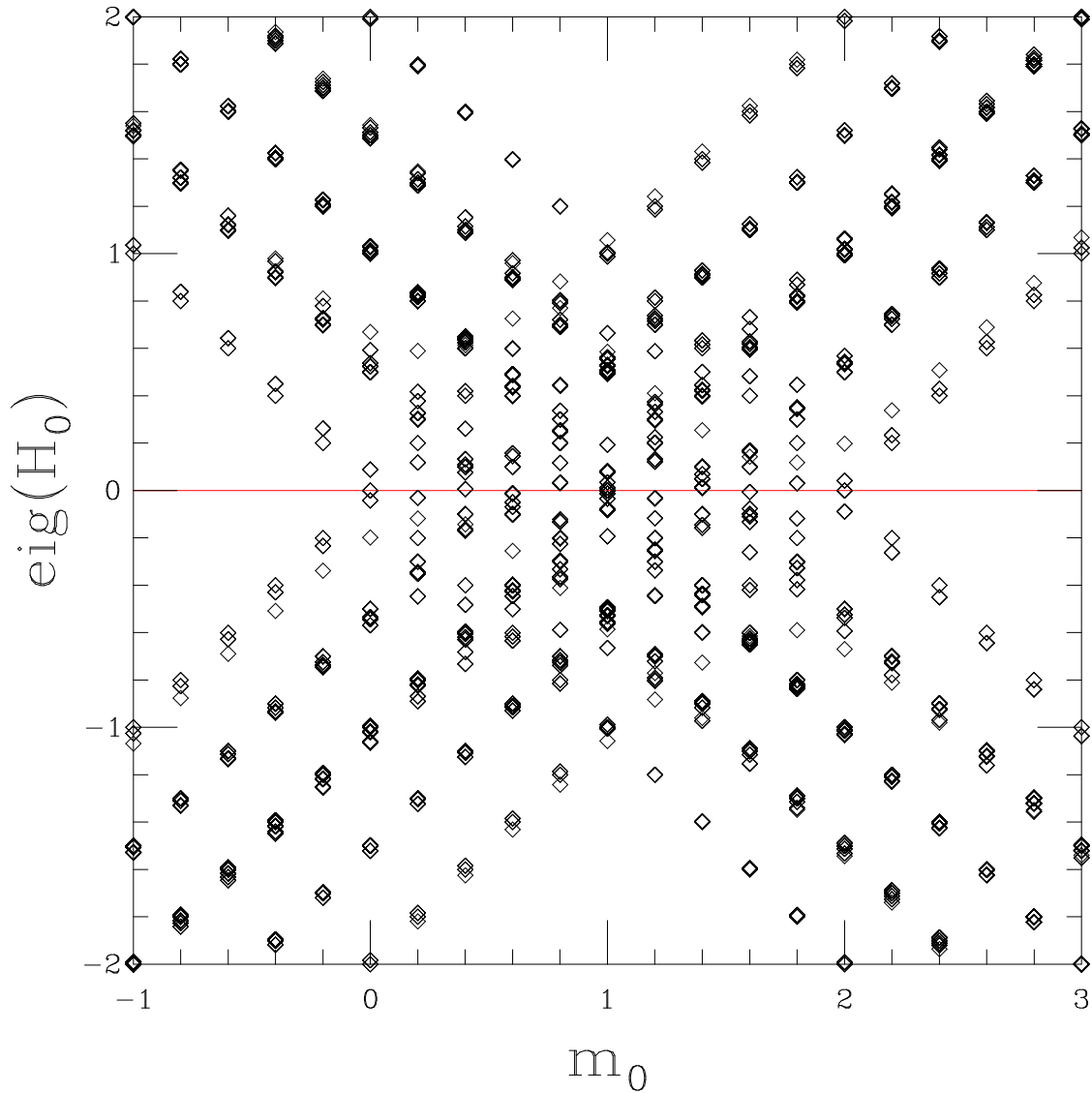}
\caption{\label{fig:eig_h_a5_0.0}The spectrum of $H_0$ \textit{vs.}\ $m_0$
  for an SU(3) \textit{``instanton''} background on a $2^4$ lattice
  of hypercubes.}
\end{figure}

For example, let us consider the spectrum of $H_0$.  Following methods identical
to \cite{Narayanan:1993wx,Narayanan:1994sk,Narayanan:1993ss, Narayanan:1995gw}
we find that
\begin{equation}
H_0 \left( \begin{array}{c} u \\ v \end{array} \right) = 0
\quad \Rightarrow \quad
u^\dagger \Delta u + v^\dagger \Delta v - 2 m_0 = 0 .
\label{eq:H_a5_0_m0_range_eq}
\end{equation}
Because the matrices $\delta(y + \hat\mu - \yp) V_\mu(y)$ are unitary
the range for which this equation can have a solution (for all Brillouin zones)
is
\begin{equation}
0 < m_0 < 2 \quad (a_5 \to 0).
\label{eq:H_a5_0_m0_range}
\end{equation}
For $m_0$ in the above range $H_0$ can have zero eigenvalues that
via the overlap formalism correspond to a change of index and to exact
and robust zeros of the fermionic determinant.  This can also be seen
graphically in Fig.~\ref{fig:eig_h_a5_0.0}. The background field configuration
is a smooth gauge field that has non-trivial topology
\cite{Narayanan:1993ss,Narayanan:1995gw}.
The plaquette value for that configuration (\textit{i.e.}\ the sum
in Eq.~(\ref{eq:action_G})) is about $0.05$.  For the rest of the paper
we will refer to this configuration
as the \textit{``instanton''} configuration. 

The crossing diagrams were done for fermions in the Saclay basis
with gauge fields $V_\mu(y)$ defined on a $2^4$ lattice of hypercubes.
Since no attempt is made here to make a connection with the topological charge
of the underlying gauge configurations $U_\mu(x)$, the reader can regard
the gauge fields as examples ``pulled out of a hat.''  All numerical analysis
was done by full diagonalization of the relevant matrices
using the LAPACK libraries and an IBM-T20 Think Pad
(which by the way performed brilliantly).

However, numerical simulations are done at non-zero $a_5$, typically
at $a_5=1$.  In standard DWF there is an exact connection
between zero eigenvalues of $H_0$ and unit eigenvalues of the transfer matrix
at any $a_5$.  This correspondence does not hold here.  As a result
the analysis is more complicated.  In other words
\begin{equation}
|\text{eig}(-T^2)| = 1 \quad \not\Rightarrow \quad \text{eig}(H_0) = 0,
\quad a_5 \ne 0.
\label{eq:H_T_non_equiv}
\end{equation}
From Eqs.~(\ref{eq:T_anticom}) and (\ref{eq:T_antiherm}) we deduce that
the spectrum of $-T^2$ is strictly real, positive and doubly
degenerate because $T$ is anti-Hermitian and anticommutes with
$\xfive$. In $2n=2$ dimensions that would be the end of the story because
there are only $2^n=2$ flavors. Two exact zero modes are produced for
every crossing in the Hamiltonian spectrum. In $2n=4$ dimensions there
are $2^n=4$ flavors and we do not have an exact correspondence between
the degeneracy of the unit magnitude eigenvalues of $T$ and the number
of flavors.  This is problematic since this is a basic and
defining property for a chiral theory.

\begin{figure}
\includegraphics[width=0.9\columnwidth]{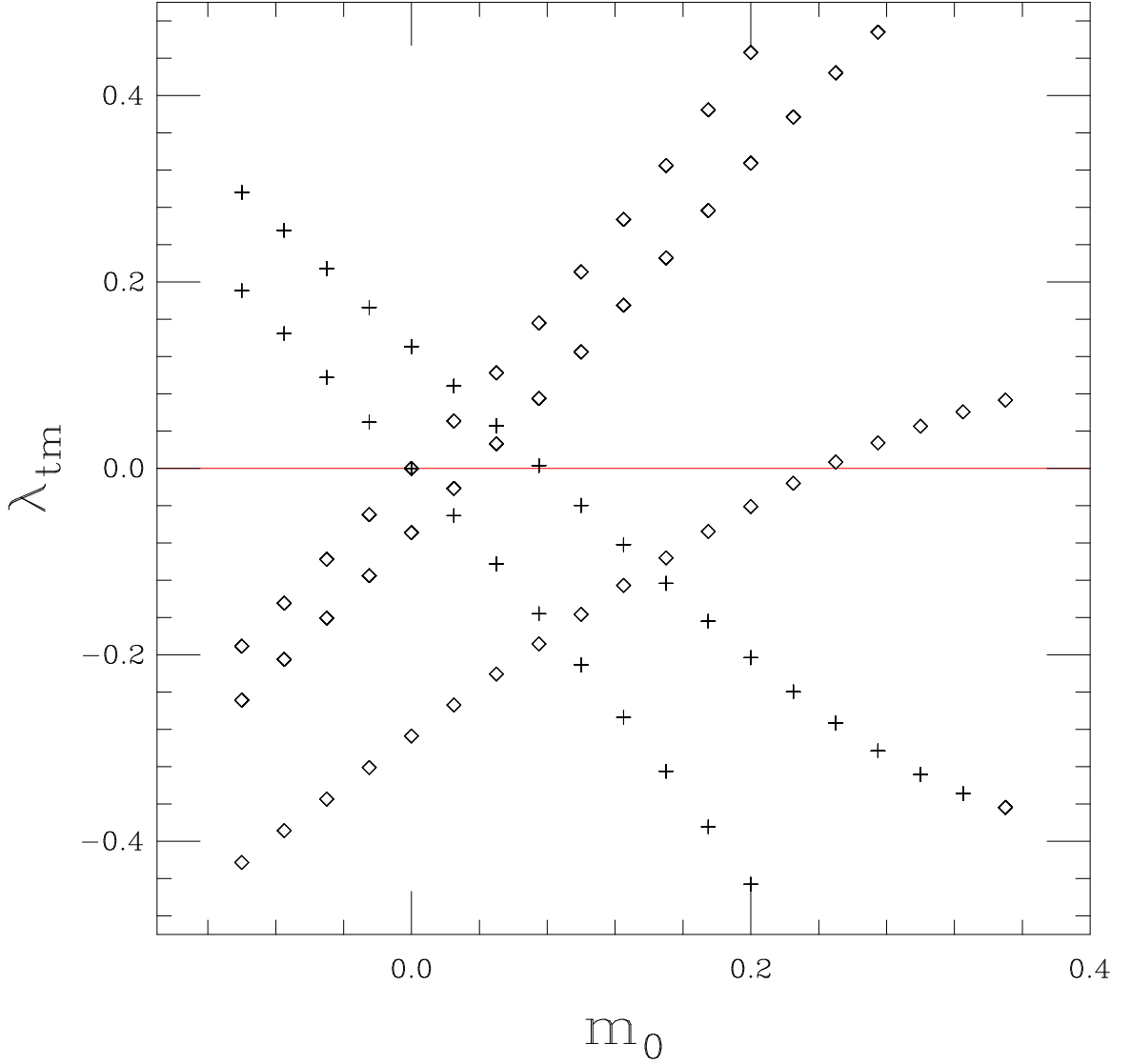}
\caption{\label{fig:eig_a5_1.00_chiral}$\lambda_{tm}$ \textit{vs.}\ $m_0$
  for $a_5=1$ and an SU(3) \textit{``instanton''} background
  (plaquette $\approx 0.05$) on a $2^4$ lattice of hypercubes.
  The diamonds represent eigenvectors of $T^2$ with chirality $+1$
  while the pluses with $-1$.  All eigenvalues are four-fold degenerate
  and are indistinguishable by the graphics.  For an example please refer
  to Table~\ref{tab:eig_a5_1.00_chiral}.}
\end{figure}

For the \textit{``instanton''} background the eigenvalue crossing diagram of 
\begin{equation}
\lambda_{tm} = \log[\text{spectrum}\ (-T^2)]
\label{eq:lambda_tm}
\end{equation}
\textit{vs.}\ $m_0$ is given in Fig.~\ref{fig:eig_a5_1.00_chiral} for $a_5=1$.
A closeup of the region around $m_0$=0 is shown and the eigenvalues are marked
depending on the sign of the $\qbar \gfive q$ where $q$ is the corresponding
eigenvector.  All eigenvalues in Fig.~\ref{fig:eig_a5_1.00_chiral} are
four-fold degenerate! The symbols for each eigenvalue are there but are
literally overlapping.  An example is given
in Table~\ref{tab:eig_a5_1.00_chiral}.  Also, the reader should observe that
in Fig.~\ref{fig:eig_a5_1.00_chiral} the transfer matrix eigenvector
chirality does follow each flow line.  This ensures that the four modes
are of the same chirality and therefore a crossing should correspond
to a net change of four in the index.  It is of fundamental importance
that this occurs here as it reassures us that on a given boundary
of the extra dimension there are four flavors of light chiral fermions
with the same chiral charge.

\begin{table}
\caption{\label{tab:eig_a5_1.00_chiral}The near zero spectrum of $\log(-T^2)$
  for $m_0=0.2$ for an SU(3) \textit{``instanton''} background
  on a $2^4$ lattice of hypercubes.}
\begin{tabular}{||l|l||} \hline
$m_0$  & $\log{\lambda(-T^2)}$ \\ \hline
0.2    & -0.0408683           \\ \hline
0.2    & -0.0408683           \\ \hline
0.2    & -0.0408683           \\ \hline
0.2    & -0.0408683           \\ \hline
0.2    &  0.327367            \\ \hline
0.2    &  0.327367            \\ \hline
0.2    &  0.327367            \\ \hline
0.2    &  0.327367            \\ \hline
\end{tabular}
\end{table}

\begin{figure}
\includegraphics[width=0.9\columnwidth]{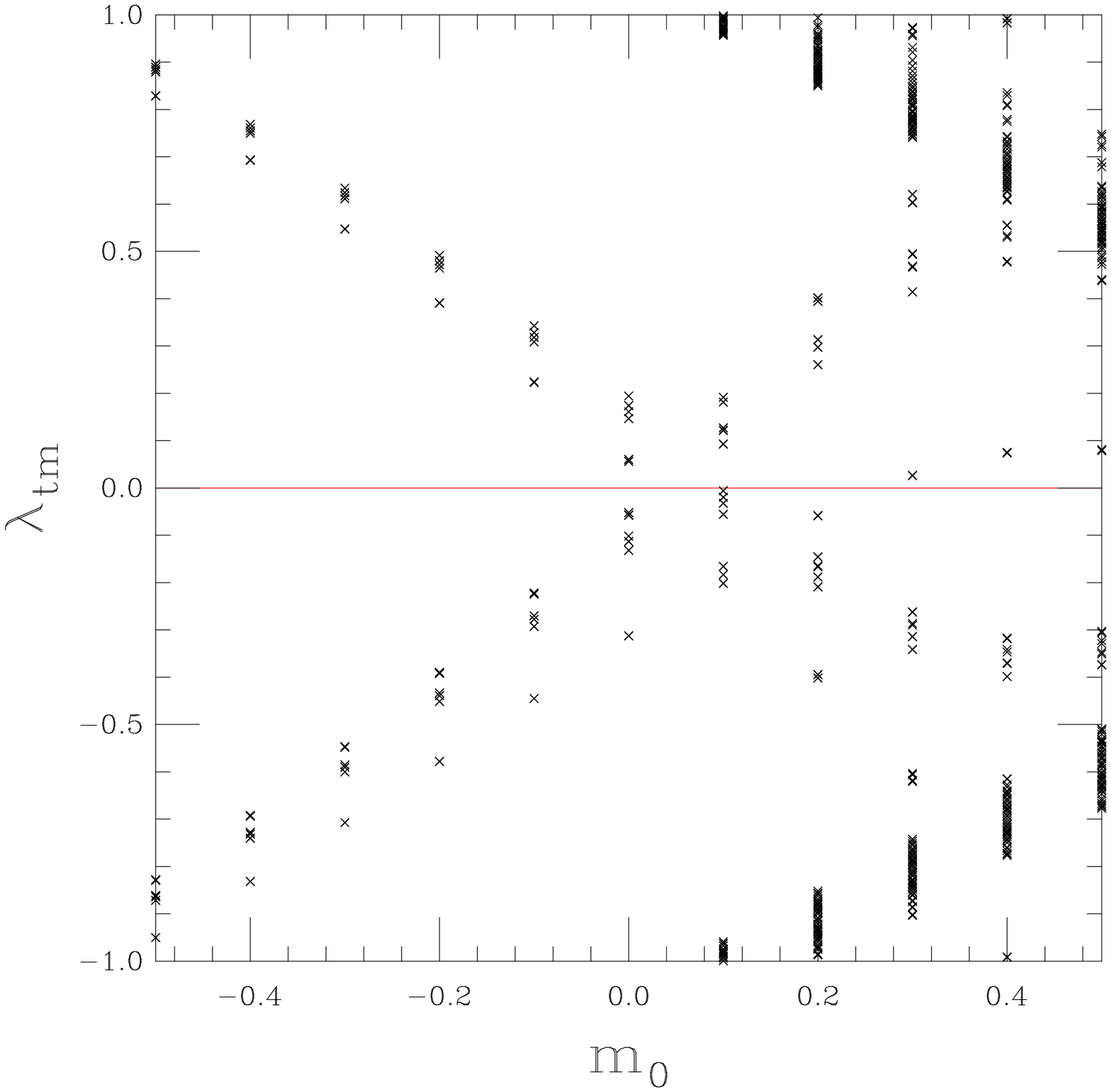}
\caption{\label{fig:eig_a5_1.00_rough}$\lambda_{tm}$ \textit{vs.}\ $m_0$
  for $a_5=1$ and a \textit{``rough''} SU(3) background
  (plaquette $\approx 0.85$) on a $2^4$ lattice of hypercubes.  All eigenvalues
  are two-fold degenerate and are indistinguishable by the graphics.
  Although there is no exact four-fold degeneracy the graphics
  can hardly distinguish the non-degeneracy.  Please refer
  to Table~\ref{tab:eig_a5_1.00_rough}.}
\end{figure}

For a very rough background gauge field (plaquette $\approx 0.85$)
the crossing diagram is given in Fig.~\ref{fig:eig_a5_1.00_rough}.
The eigenvalues are doubly degenerate and very nearly but not quite
four-fold degenerate. The non degeneracy is small and almost non visible.
One of the worst cases is presented in Table~\ref{tab:eig_a5_1.00_rough}.
The lack of exact four-fold degeneracy should be the subject
of further research but it is obviously small even
on this extreme background gauge field that is unlikely to occur
in current numerical simulations of QCD.  Furthermore, it would be interesting
to study the transfer matrix in $2n \ge 6$ dimensions, where the required
near-degeneracy should be $2^n$, to see if the same behavior
persists.

\begin{table}
\caption{\label{tab:eig_a5_1.00_rough}The near zero spectrum of $\log(-T^2)$
  for $m_0=0.3$ for a \textit{``rough''} SU(3) background
  (plaquette $\approx 0.85$) on a $2^4$ lattice of hypercubes.}
\begin{tabular}{||l|l||}        \hline
$m_0$   & $log{\lambda(-T^2)}$ \\ \hline
0.3	&	-0.290045    \\ \hline
0.3	&	-0.290045    \\ \hline
0.3	&	-0.286291    \\ \hline
0.3	&	-0.286291    \\ \hline
0.3	&	-0.262638    \\ \hline
0.3	&	-0.262638    \\ \hline
0.3	&	-0.261551    \\ \hline
0.3	&	-0.261551    \\ \hline
0.3	&	0.0260566    \\ \hline
0.3	&	0.0260566    \\ \hline
0.3	&	0.0272779    \\ \hline
0.3	&	0.0272779    \\ \hline
0.3	&	0.414225     \\ \hline
0.3	&	0.414225     \\ \hline
0.3	&	0.414597     \\ \hline
0.3	&	0.414597     \\ \hline
\end{tabular}
\end{table}

Even with the lack of exact four-fold degeneracy, if we choose $m_0$ away
from the crossing region then the fermion determinant will still have
four exact zeros.  This will only break down if $m_0$ is chosen between the two
nearly degenerate sets of crossings. Of course, based on DWF studies
at lattice spacings used in today's simulations one
expects dense crossings in the usable range of $m_0$. Then there
will be configurations for which $m_0$ is between double-crossings
that have split. As far as topology is concerned this will break
flavor to some degree. Nevertheless, since in DWF the dense crossings
correspond to small instantons and are unphysical we would expect that
as the lattice spacing becomes smaller such configurations will
become less important.

\begin{figure}
\includegraphics[width=0.9\columnwidth]{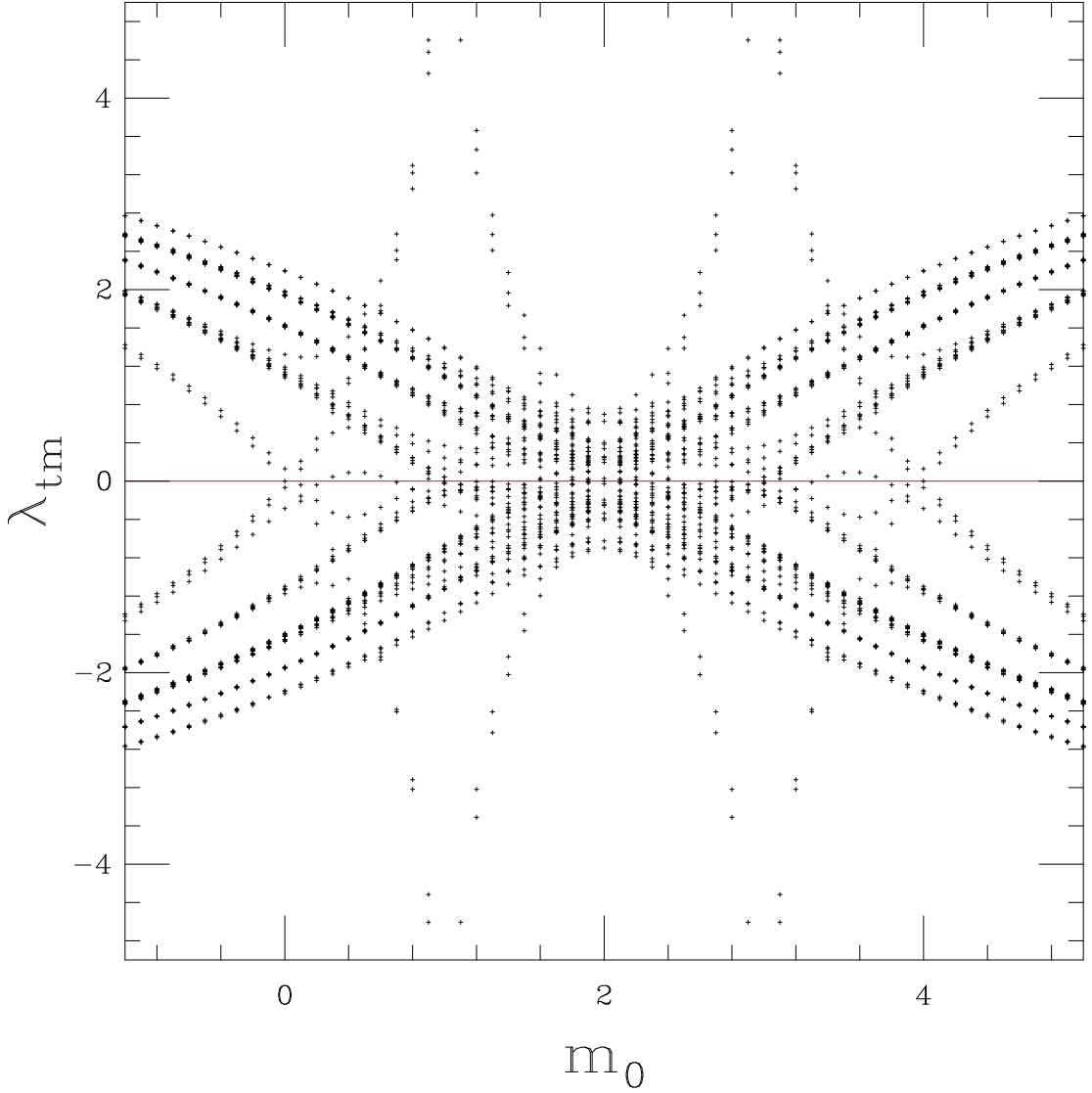}
\caption{\label{fig:eig_a5_1.00}$\lambda_{tm}$ \textit{vs.}\ $m_0$ for $a_5=1$
  and an SU(3) \textit{``instanton''} background  (plaquette $\approx 0.05$)
  on a $2^4$ lattice of hypercubes.}
\end{figure}

\begin{figure}
\includegraphics[width=0.9\columnwidth]{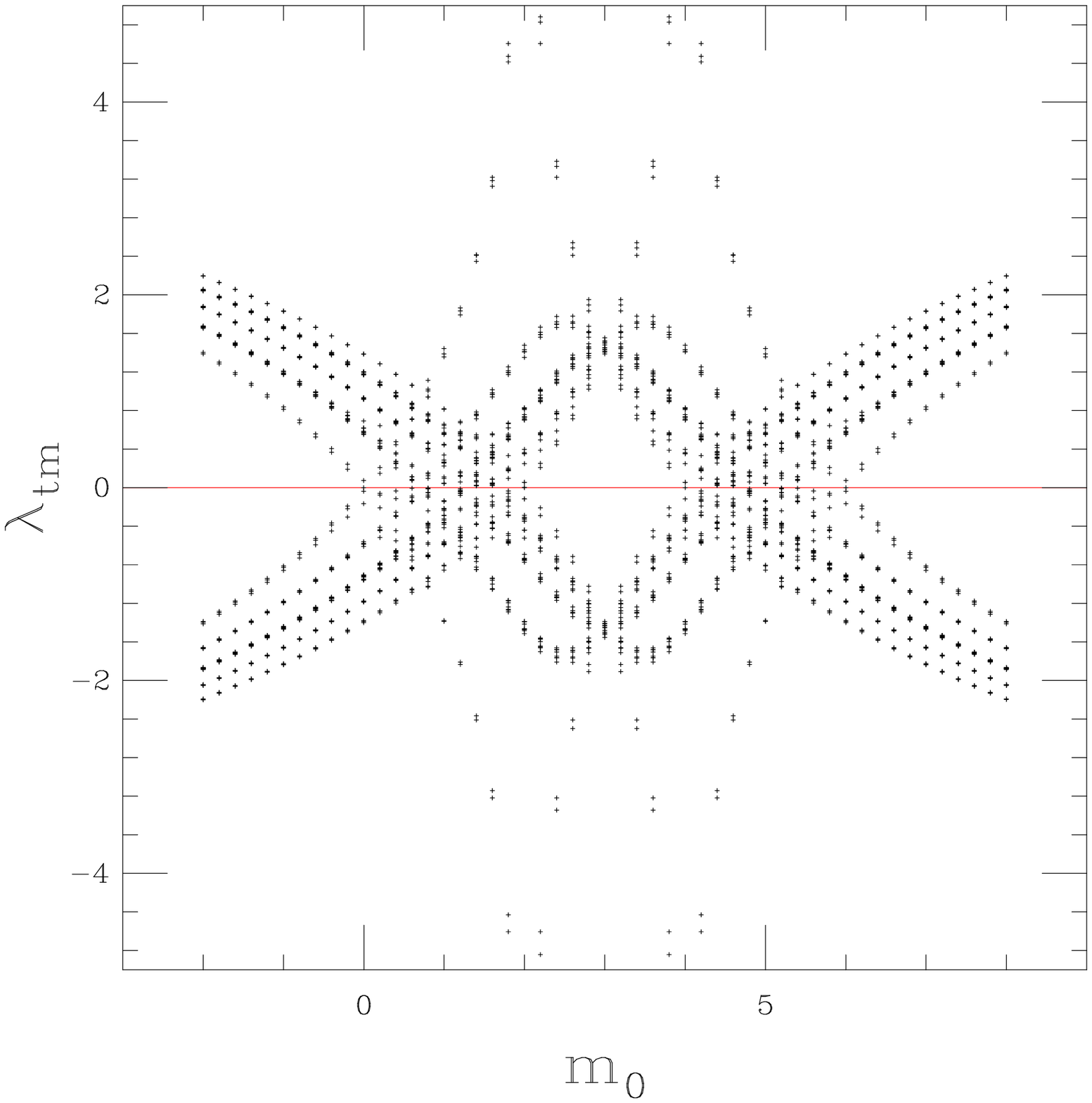}
\caption{\label{fig:eig_a5_0.50}$\lambda_{tm}$ \textit{vs.}\ $m_0$ for $a_5=0.5$
  and an SU(3) \textit{``instanton''} background  (plaquette $\approx 0.05$)
  on a $2^4$ lattice of hypercubes.}
\end{figure}

\begin{figure}
\includegraphics[width=0.9\columnwidth]{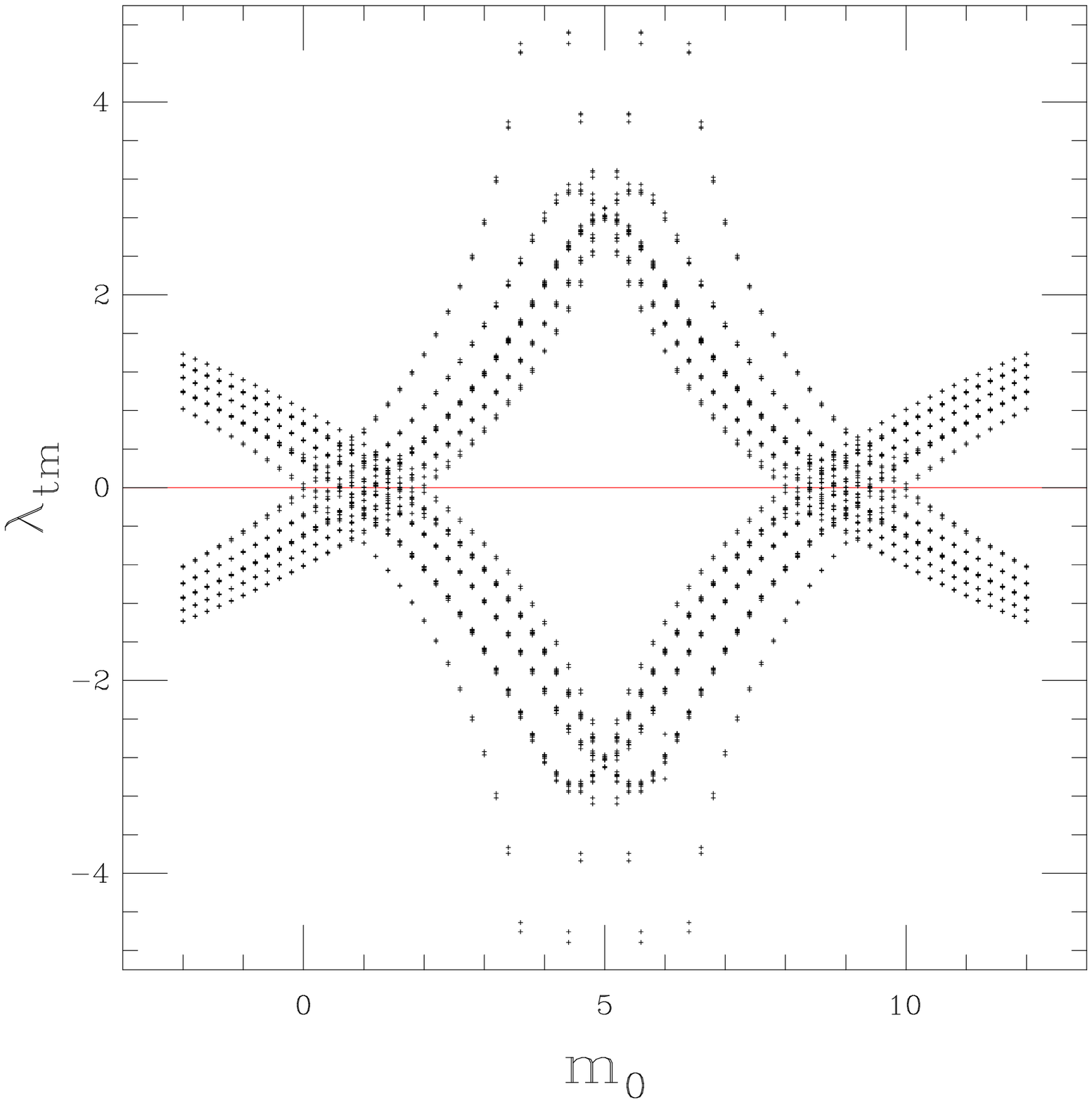}
\caption{\label{fig:eig_a5_0.25}$\lambda_{tm}$ \textit{vs.}\ $m_0$
  for $a_5=0.25$ and an SU(3) \textit{``instanton''} background
  (plaquette $\approx 0.05$) on a $2^4$ lattice of hypercubes.}
\end{figure}

Nevertheless it is instructive to further study the spectrum of $T$ for
unit magnitude eigenvalue. Because $T$ contains the inverse of $B$  
it is hard to study analytically. However, in the subspace 
$|\lambda_{tm}| = 1$ we can proceed as in \cite{Narayanan:1993wx,
Narayanan:1994sk,Narayanan:1993ss,Narayanan:1995gw}.
In particular we find that
\begin{equation}
T \left( \begin{array}{c} u \\ v \end{array} \right)
= \pm\ i\ T \left( \begin{array}{c} u \\ v \end{array} \right)
\quad \Rightarrow \quad
H_p \left( \begin{array}{c} u \\ v \end{array} \right)
= 0
\label{eq:T_Hp}
\end{equation}
where
\begin{equation}
H_p =  \left( \begin{array}{cc}
    1 + a_5 i B   &  a_5 C \\
    a_5 C^\dagger        & -1 - a_5 i B
\end{array} \right) .
\label{eq:Hp}
\end{equation}
So, we can study the crossings of the spectrum
of this \textit{pseudo}-Hamiltonian $H_p$. The crossing range can be determined
as before by using the unitarity of the matrices
$\delta(y + \hat\mu - \yp) V_\mu(y)$.  There are two crossing ranges
\begin{equation}
0 < m_0 < 2 \quad \text{and} \quad \frac{2}{a_5} < m_0 < \frac{2}{a_5} + 2 .
\label{eq:m0_cross_any a5}
\end{equation}
For the \textit{``instanton''} background the crossing diagrams
(for all Brillouin zones) are shown for $a_5=1$
in Fig.~\ref{fig:eig_a5_1.00}, $a_5=0.5$ in Fig.~\ref{fig:eig_a5_0.50}
and $a_5=0.25$ in Fig.~\ref{fig:eig_a5_0.25} and the reader can see
the agreement with Eq.~(\ref{eq:m0_cross_any a5}).

But this is not all.  We had a hard time at first because we used
SU(2) gauge fields.  And the spectrum was always inexplicably not two-fold
but four-fold degenerate, for \textit{any} gauge field (smooth or rough)!
This is the second surprise. And it will not be investigated further here.
After all, what good is a paper that does not leave some mystery behind...
In any case this property is likely linked to the fact that $-1$ is part
of the groups SU(2), SU(4), \dots but is not part of the group SU(3),
\textit{etc.}

\section{The single component basis and simulating SDWF}
\label{sec:single_comp}

In the previous sections we discussed the properties of the SDWF Dirac operator
in the basis proposed by the Saclay group.  By now, the reader has seen
the advantage of this approach in determining the properties of robust zeros
of the transfer matrix.  In the past, technical problems have disfavored
direct simulation of dynamical fermions in this basis in lieu
of the simpler single component basis, where there is just
a single fermionic spin-flavor degree of freedom per site.

In this section, we will first discuss the construction of the SDWF
in the single component basis with an emphasis on preserving,
to the maximum extent possible, all of the symmetries
of section~\ref{sec:symmetries}.  One consequence is that
the spin-flavor algebra, \textit{i.e.}\ $\gfive^2 = \gxone$, will be broken
and only recovered in the continuum limit.  Next, we will discuss a technique
where the spin-flavor algebra is restored at the expense
of some lattice symmetries.  We believe these lattice symmetries
can be restored by proper stochastic averaging.  Finally, we will discuss
a new algorithm based on \textit{double regularization} for simulating
staggered (and SDWF) fermions directly in the Saclay basis.  As is typically
the case in our field, performance during numerical simulation of QCD
will likely determine which of the three proposals survive.
We feel that further research in this area is needed.

\subsection{Single component basis without projection}
\label{sub:single_comp_wo_proj}

Let us first review the staggered action in the single component basis
and the transformation connecting it to the Saclay basis.
From Eqs.~(\ref{eq:D_F}) through (\ref{eq:deriv_4}), with $m_0 = 1/a_5$,
the equivalent $2n$ dimensional Dirac operator is usually written
\begin{equation}
D(x,x^\prime) \chi(x^\prime) = \frac{1}{2} \sum_\mu
(-1)^{\hat\mu \cdot \eta(x)} \left[
  U_\mu(x) \delta(x+\hat\mu-x^\prime)
  - U_\mu^\dagger(x) \delta(x-\hat\mu-x^\prime) 
\right] \chi(x^\prime)
\label{eq:D_stag_sc}
\end{equation}
with the components of the binary vector $\eta$ given by
\begin{eqnarray}
\eta_1(x) & = & 0 , \nonumber \\
\eta_2(x) & = & x_1 \pmod 2 , \nonumber \\
& \vdots & \nonumber \\
\eta_{2n}(x) & = & x_1 + \cdots + x_{2n-1} \pmod 2 .
\label{eq:eta}
\end{eqnarray}
In the free theory,
the unitary transformation from single component site-wise fields $\chi(x)$
to the hypercubic fields $\psi(y)$ is simple.  As in section~\ref{sec:sdwf},
if we label sites on the hypercube starting at the origin $O$
by a binary vector $A$ the transformation is
\begin{equation}
\psi_{\alpha a}(y) \propto \Gamma_{\alpha a, A} \chi(2y+O+A)
\end{equation}
with the rows of the $2^{2n}\times 2^{2n}$ dimensional matrix $\Gamma$
indexed by the various combinations of spin $\alpha$ and flavor $a$ indices
and the columns indexed by the corners of the hypercube $y$.  Specifically,
the components of the free $\Gamma$ may be chosen as
\begin{equation}
\Gamma_{\alpha a, A} = \left[
\gamma_1^{A_1} \gamma_2^{A_2} \times \cdots \times \gamma_{2n}^{A_{2n}}
\right]_{\alpha a}
\end{equation}
where the row index $\alpha$ and column index $a$ of the $2^n$ dimensional
representation of the Clifford algebra are interpreted as spin and flavor
indices.

When the staggered action is made gauge invariant, differences will arise
between the two formulations.  In the single component basis, fermions are
only coupled to nearest neighbor sites, so simple links are all that is needed
to preserve gauge invariance.  Of course, longer paths could be used to link
sites and indeed are often used to implement an improvement program.  In the
Saclay basis, since fermion fields are associated with hypercubes, the gauge
fields must be used to move the components on the hypercube to some
common point where the hypercubic field can be assembled.  While it may seem
that the resulting actions could be completely different, they have the same
terms, site by site, that differ only in the choice of paths used to connect
nearest neighbors.  Hence, they have the same continuum limit.

However, this is not the end of the story.  For example, when moving
a component two sites on the hypercube for the construction
of the hypercubic field there are two equivalent minimum distance paths
from which to choose.  Choosing one path over the other will preserve
the unitarity of $\Gamma$ but break the rotation by $\pi/2$ symmetry.
Conversely, choosing to average over both paths
preserves rotations but means $\Gamma$ need not be unitary,
and thus potentially singular for sufficiently rough gauge fields.
For SDWF, some terms in the action may also break
the shift by one lattice spacing symmetry.  The well known source
of the problem is the imposition of the artificial hypercubic structure
for the identification of spin and flavor degrees of freedom
\cite{Mitra:1983bi}.

Thus, a conservative approach is to abandon a transcription from hypercubic
bases and directly use the single component basis
and the technique of Golterman and Smit \cite{Golterman:1984cy}.
Using the symmetric shift operator
with the binary vector $\zeta(x)$ and scalar $\varepsilon(x)$
\begin{eqnarray}
& E_\mu(x,x^\prime) = (-1)^{\hat\mu \cdot \zeta(x)} \frac{1}{2} \left[
    U_\mu(x) \delta(x + \hat\mu - x^\prime)
    + U_\mu^\dagger(x-\hat\mu) \delta(x - \hat\mu - x^\prime)
  \right] & \\
& \zeta_1(x) = x_2 + \cdots + x_{2n} \pmod 2,\ \cdots,
  \ \zeta_{2n-1} = x_{2n} \pmod 2,\ \zeta_{2n} = 0 & \\
& \varepsilon(x) = x_1 + \cdots +x_{2n} \pmod 2 &
\end{eqnarray}
we write the $m_0$ term
\begin{equation}
\frac{1}{2}\left( \frac{1}{a_5} - m_0 \right)
\sum_\mu \left( \gamma_5 \otimes \xi_{5\mu} \right) \psi(y) \to
\frac{1}{2}\left( \frac{1}{a_5} - m_0 \right) (-1)^{\varepsilon(x)}
\sum_\mu E_\mu(x,x^\prime) \chi(x^\prime)
\label{eq:GS_m0_term}
\end{equation}
and the chiral projection operators are constructed using
\begin{equation}
\left(\gamma_5 \otimes \openone\right) \psi(y) \to
\frac{(-1)^{\varepsilon(x)}}{(2n)!} \sum_{\mu_1 \cdots \mu_{2n}}
\epsilon_{\mu_1 \cdots \mu_{2n}} E_{\mu_1}(x,x^{(1)}) \times \cdots
\times E_{\mu_{2n}}(x^{(2n-1)}, x^{(2n)}) \chi(x^{(2n)})
\label{eq:GS_gamma_5}
\end{equation}
where $\epsilon_{\mu_1 \cdots \mu_{2n}}$ is the totally antisymmetric tensor
and the summation over the $2n$ site vectors $x^{(1)}, \cdots, x^{(2n)}$
is implied.  The upside to this approach is that it preserves the staggered
symmetries to the extent possible.  The downside is that chiral projection
is no longer exact, except in the continuum limit.  Of course, this is
a different manifestation of the same problem that makes the transformation
to the Saclay basis non-unitary, where projection is exact.

\subsection{Exact projection with stochastic symmetrization}
\label{sub:exact_proj_w_stoch_sym}

The second approach addresses the projection problem at the expense
of breaking some symmetries, which can be restored in the ensemble average
as described below\footnote{GTF would like to thank M.\ Di Pierro for
a useful discussion on this point.}.  Our example will use the hypercubic
basis of Daniel and Sheard \cite{Daniel:1988aa} but equivalent examples
are to construct a unitary transformation $\Gamma$ into the Saclay basis
or even to restrict the Golterman--Smit operators to single paths between
sites.

Quickly reviewing the Daniel--Sheard formulation, we want to construct
\textit{``local''} fermion bilinears of definite spin and flavor, where local
means local to the hypercube, from the single component states $\chi(x)$.
We identify the hypercubic Daniel--Sheard fields by a simple relabeling:
$\chi_A(y) = \chi(x)$ and $x=2y+O+A$ as before. Local bilinears are written
\begin{equation}
\overline\chi_A(y) (\overline{\gamma_S \otimes \xi_F})_{AB}
\chi_B(y) = \sum_{x,x^\prime} (-1)^{\phi(x,x^\prime)} \overline\chi(x)
\mathcal{U}(x,x^\prime) \chi(x^\prime)
\end{equation}
where $x,x^\prime$ are summed over the hypercube and $\mathcal{U}(x,x^\prime)$
represents the links chosen to make the bilinear gauge invariant.
The notation is $\gamma_S = \gamma_1^{S_1} \times \cdots \times
\gamma_{2n}^{S_{2n}}$ and the phase factor is computed from
\begin{equation}
(\overline{\gamma_S \otimes \xi_F})_{AB} \ \Rightarrow
\ \phi(A,B) = \frac{1}{2^n} \text{Tr} \left(
\gamma_A^\dagger \gamma_S \gamma_B \gamma_F^\dagger
\right).
\end{equation}
As an aside, this gives exactly the same terms appearing
in Eqs.~(\ref{eq:GS_m0_term}) and (\ref{eq:GS_gamma_5}) provided you keep
only the terms on a single hypercube.

As mentioned before, imposing a hypercubic structure introduces problems
with maintaining the staggered symmetries for arbitrary spin and flavor
choices.  Our proposal is at the beginning of each update step
of whatever update algorithm, first choose the origin $O$ at random
from the $2^{2n}$ ways of imposing the hypercubic structure on the lattice.
Next, choose at random only one of the minimum distance paths on the hypercube
for making bilinears gauge invariant with the restriction that the same path
is used in both directions.  Thus, $\mathcal{U}(A,B)$ is unitary and
$\mathcal{U}^\dagger(A,B) = \mathcal{U}(B,A)$.  Note that different paths
can be used on different hypercubes.  Choosing a random hypercubic structure
and random paths on the hypercubes at each update step ensures that
symmetry breaking effects due to these choices will cancel out
in the ensemble average.  The purpose of choosing only one path per pair
of corners on the hypercube is to guarantee the chiral projection property.
For example, $(\overline{\gamma_5 \otimes \openone})^\dagger
\left[ (\overline{\gamma_5 \otimes \openone}) \chi \right] \to \chi$
which is the same as the continuum where we normally choose Hermitian gamma
matrices.

\subsection{Doubly regularized staggered fermions}
\label{sub:doub_reg_stag_ferm}

The third proposal is specific to the Saclay basis but applies equally well
to SDWF and staggered fermions with Pauli--Villars fields.  Following the
second proposal, we can construct at each update step a unitary transformation
from the single component basis to the Saclay basis that will certainly depend
on the gauge field but not on the value of the mass $m_f$.  Since the
fermionic action and the Pauli--Villars action only differ by the value
of $m_f$, then the contributions from the transformation will cancel between
fermions and the pseudofermions.  Specifically, the fermionic partition
function on a fixed gauge background and for finite lattice spacing,
volume and $L_s$ is
\begin{eqnarray}
Z_{\text{F}}\left[U\right] & = &
\int \left[ d\overline\chi_{\text{F}} d\chi_{\text{F}} \right]
\int \left[ d\phi_{\text{PV}}^\dagger d\phi_{\text{PV}} \right] 
e^{\overline\chi_{\text{F}} \Gamma^\dagger D_{\text{F}}(m_f)
  \Gamma \chi_{\text{F}} - \phi_{\text{PV}}^\dagger \Gamma^\dagger
  D_{\text{F}}(m_f=1) \Gamma \phi_{\text{PV}} } \nonumber \\
& = & \frac{\det\Gamma^\dagger \ \det D_{\text{F}}(m_f) \ \det\Gamma}{
\det\Gamma^\dagger \ \det D_{\text{F}}(m_f=1) \ \det\Gamma} =
\frac{\det D_{\text{F}}(m_f)}{\det D_{\text{F}}(m_f=1)} \nonumber \\
& = & \int \left[ d\overline\Psi d\Psi \right]
\int \left[ d\Phi^\dagger d\Phi \right] e^{\overline\Psi D_{\text{F}}(m_f)
  \Psi - \Phi^\dagger D_{\text{F}}(m_f=1) \Phi}
\end{eqnarray}
which is what we had back in Eq.~(\ref{eq:partition_func}).  In practice,
we do not need to specify the paths chosen for the basis transformation
since they cancel from the path integral.  But, it would still be important
to choose at random the origin $O$ of the hypercubic structure
at each update step to avoid violations of the shift
by one lattice spacing symmetry.  We would like to emphasize again
that this proposal should work for staggered fermions
with added Pauli--Villars fields and we believe this is another example
of the potential of \textit{double regularization} to improve the usefulness
of existing fermion actions by canceling lattice artifacts
\cite{Fleming:2000bk}.  Also, notice that the number of fermionic degrees
of freedom is the same as in the single component basis because the Saclay
fields are defined on hypercubes.

Of course, some of the ideas presented in this section for implementing SDWF
for numerical simulation have been discussed before, \textit{e.g.}\ the idea
for stochastic restoration of staggered symmetries is certainly descended
from the work of Christ, Freidberg and Lee \cite{Christ:1982zq}.

\section{Alternative actions}
\label{sec:alt}

The SDWF action considered here is not unique. It is possible that actions
with better scaling properties may be constructed using improved fields
in the same spirit as with staggered fermions (see \cite{Luo:1997vt,Lee:1999zx}
and references therein).  Additionally, in our earlier work
\cite{Fleming:2001ua} we introduced the domain wall defect using a local
mass term (distance zero) which preserved the shift by one lattice spacing
symmetry (among others) and broke the U(1)$\times$U(1) chiral symmetry.
In this work, we considered a distance one mass term which preserves
the chiral symmetry and breaks the shift symmetry. We view this
as a better choice because the additive renormalization it produces
\cite{Mitra:1983bi} merely contributes to the flavor breaking term
that our domain wall formulation is designed to eliminate.  It is possible
that other distance mass terms might prove useful in the future
and even have a faster exponential rate of restoration
of flavor symmetry.  We emphasize that the primary requirements
for these mass terms are that they be of the order of the cutoff
and commute with the operators in Eq.~(\ref{eq:zero_mode_problem}).

\section{Conclusions}
\label{sec:conclusions}

In this paper a different lattice fermion regulator was presented.
Staggered domain wall fermions are defined in $2n+1$ dimensions and describe
$2^n$ flavors of light lattice fermions with exact U(1)$\times$U(1)
chiral symmetry in $2n$ dimensions.  The full SU($2^n$)$\times$SU($2^n$)
flavor symmetry is recovered as the size of the extra dimension is increased.
SDWF give a different perspective into the inherent flavor mixing of lattice
fermions and by design present an advantage for numerical simulations
of lattice QCD thermodynamics.  We have paid particular attention
to the chiral and topological index properties of the SDWF Dirac operator
and its associated transfer matrix.  In the limit where the lattice spacing
in the extra dimension $a_5$ tends to zero the corresponding Hamiltonian
$H_0$ is proportional to the identity in flavor space illustrating
the complete absence of flavor mixing.

For a semi-infinite extent in the extra dimension, the theory has four
chiral fermions with the same chiral charges and is anomalous.
To construct an anomaly free theory, we must use such ``quadruplets''
with charges as dictated by the corresponding anomaly cancellation condition.
This is completely analogous to the case of Wilson DWF.

However, there are still a number of unresolved issues related
to this formulation which need to be studied in future work.
In particular:

\begin{enumerate}

\item[1] For QCD, the nearly four-fold crossing degeneracy
         of the Hamiltonian must be investigated thoroughly.

\item[2] SDWF should be implemented for numerical simulation
         according to the proposals of section~\ref{sec:single_comp}.
         In the broken phase of QCD, it is obviously important to confirm
         that one pion is a pseudo-Goldstone boson and that
         the remaining fourteen non-singlet pions become degenerate
         with the pseudo-Goldstone boson as $L_s \to \infty$.  Also,
         the expected robustness of topological zero modes should be confirmed
         as was done for DWF \cite{Chen:1998ne}.

\item[3] We have presented an analysis of the zeros of the SDWF Hamiltonian
         through the pseudo-Hamiltonian $H_p$ and in the limit $a_5 \to 0$
         where the flavor breaking is trivially absent in $H_0$.  A derivation
         and analysis of the full spectrum of the Hamiltonian for general $a_5$
         is needed.

\item[4] Since the nearly degenerate four-fold crossings in the spectrum
         of the Hamiltonian have the same chiral charge, the conserved currents
         of the full SU($2^n$)$\times$SU($2^n$) symmetry must exist and can be
         constructed in the overlap formalism \cite{Narayanan:1995gw}.
         Simpler constructions of these currents may be possible.  Constructing
         and measuring the conservation of these currents in simulations
         is important.

\item[5] SDWF were constructed with the simulation of QCD thermodynamics
         in mind because of the importance of having a continuous subgroup
         of chiral symmetry for any $L_s$.  It is worth confirming that
         this gives SDWF some advantage over DWF in looking
         for critical fluctuations
         at the finite temperature QCD phase transition.

\item[6] It would be very interesting to add the Kogut-Sinclair
         four-fermion interaction \cite{Kogut:1997mj,Kogut:1998rg}
         to the SDWF action to enable simulation at zero quark mass
         in the region of the QCD phase transition.

\item[7] The domain wall mass term must be of the order of the lattice spacing
         and will introduce a hard breaking of some part
         of the staggered symmetry group, in our case the shift
         by one lattice spacing symmetry, causing quantum corrections
         to the SDWF transfer matrix \cite{Mitra:1983bi}.  Our analysis
         of the transfer matrix spectrum indicates a range of $m_0$ values
         can still be found where the transfer matrix behaves correctly
         in the presence of gauge field topology, as was the case
         with Wilson DWF.  We believe this issue should be studied
         thoroughly\footnote{We would like to thank M.~F.~L.~Golterman
         for useful discussions on this point.}.

\item[8] Can SDWF reveal (or has it already revealed) some new insight
         into the nature of inherent flavor mixing of lattice fermions?

\end{enumerate}

On the strength of the results of our transfer matrix analysis,
we believe that SDWF may be an attractive alternative to DWF.
The formulation is now sufficiently mature that the issues above
should now be addressed in the context of QCD.  As has been the case
with staggered and Wilson fermions in the past, we should have a choice
between SDWF and DWF for a given problem according to our resources
and preference in the near future.

\section*{Acknowledgments}

We would like to thank J.\ B.\ Kogut for continued encouragement and support
throughout the course of this project.  We would also like to thank
M.\ Di Pierro, M.\ F.\ L.\ Golterman, Y.\ Shamir and S.\ R.\ Sharpe
for useful discussions.  G.\ T.\ Fleming would like to thank the Institute
for Nuclear Theory at the University of Washington for hospitality and support
provided for part of this work.

\bibliography{paper}

\end{document}